\documentclass[preprintnumbers, floatfix, showkeys, preprintnumbers, letterpaper, twocolumn, superscriptaddress,nofootinbib]{revtex4-2}
\pdfoutput=1 
\usepackage{graphicx}
\usepackage{microtype}
\usepackage{amsmath}
\usepackage{amssymb}
\usepackage{subfigure}
\usepackage{url}
\usepackage{hyperref}
\usepackage{mathtools}
\usepackage{orcidlink}
\usepackage{slashbox}

\usepackage{xcolor}
\usepackage{color}
\usepackage{mathrsfs}
\usepackage{calrsfs}
\usepackage{amsfonts}
\usepackage{tabularx}
\usepackage{latexsym}
\usepackage{ragged2e}
\usepackage{epsfig}
\usepackage{textcomp}
\usepackage{float}

\usepackage{caption}
\DeclareCaptionJustification{justified}{\leftskip=0pt \rightskip=0pt \parfillskip=0pt plus 1fil}
\definecolor{vividviolet}{rgb}{0.62, 0.0, 1.0}
\definecolor{amaranth}{rgb}{0.9, 0.17, 0.31}
\definecolor{palatinateblue}{rgb}{0.15, 0.23, 0.89}
\definecolor{brightpink}{rgb}{1.0, 0.0, 0.5}
\definecolor{cornflowerblue}{rgb}{0.39, 0.58, 0.93}
\definecolor{deepcarminepink}{rgb}{0.94, 0.19, 0.22}
\definecolor{radicalred}{rgb}{1.0, 0.21, 0.37}

\hypersetup{ linktoc=all,
	colorlinks, linkcolor={palatinateblue},
	citecolor={brightpink}, urlcolor={amaranth}
}

\graphicspath{{Images/}}

\renewcommand{\d}[1]{\ensuremath{\operatorname{d}\!{#1}}}

\def\sideremark#1{\ifvmode\leavevmode\fi\vadjust{\vbox to0pt{\vss% the remark
			\hbox to 0pt{\hskip\hsize\hskip1em%                          will appear only
				\vbox{\hsize1.3cm\tiny\raggedright\pretolerance10000%          on the side
					\noindent #1\hfill}\hss}\vbox to8pt{\vfil}\vss}}}%
\def\beq{\begin{equation}}
\def\eeq{\end{equation}}

\begin{document}
\title{GUP Effective Metric Without GUP:\\ Implications for the Sign of GUP Parameter and Quantum Bounce}

\author{Yen Chin \surname{Ong}\orcidlink{0000-0002-3944-1693}}
\email{ycong@yzu.edu.cn}
\affiliation{Center for Gravitation and Cosmology, College of Physical Science and Technology,\\ Yangzhou University,
 Yangzhou, 225002, China}% \\180 Siwangting Road, Yangzhou City, Jiangsu Province  225002, China}
\affiliation{Shanghai Frontier Science Center for Gravitational Wave Detection, School of Aeronautics and Astronautics,\\ Shanghai Jiao Tong University, Shanghai 200240, China}

\begin{abstract}
The standard form of generalized uncertainty principle (GUP) predicts that the Hawking temperature is modified near the Planck scale and that the Bekenstein-Hawking entropy receives a logarithmic correction, consistent with other approaches to quantum gravity. However, due to the heuristic arguments in most GUP literature, it is not clear how to obtain the Schwarzschild metric that incorporates GUP correction. In this work, we try a different approach. We will start with the entropy expression with the standard logarithmic correction term, and use the recently proposed ``generalized entropy and varying-$G$ correspondence'' (GEVAG) to obtain the associated metric. We show that the Hawking temperature obtained from this metric matches the GUP version. In this sense, we have derived in a consistent and reliable manner, a metric tensor that can describe the standard GUP physics, and use it to clarify some shortcomings in the heuristic GUP approach itself. In particular, if the strict Bekenstein bound is imposed, then the GUP parameter is negative. We also speculate on the possibility that instead of a stable remnant, the final stage of black hole evaporation could be a ``bounce'' due to an effective gravitational repulsion, once higher order corrections are included.
\end{abstract}

\maketitle

\section{Introduction: Generalized Uncertainty Principle and Its Effective Metric}
Near the Planck scale, due to quantum gravity (QG) correction, we expect that the Heisenberg uncertainty principle would receive some corrections.
The generalized uncertainty principle (GUP) is one such approach that tries to incorporate QG correction phenomenologically. 
The most basic form (the ``standard form'') of GUP is\footnote{The notation $\Delta x^n$ is a short hand for $(\Delta x)^n$. Similarly for $\Delta p^n$.}
\begin{equation}\label{GUP}
\Delta x \Delta p \geqslant \frac{1}{2}\left(\hbar + \alpha L_p^2 \frac{\Delta p^2}{\hbar}\right),
\end{equation}
where $\alpha$ is a dimensionless parameter and $L_p$ the Planck length. There have been many generalizations made in the literature, but we will just focus on this basic form. A heuristic argument that makes use of the relation $\Delta x \sim r_h$ and $\Delta p \sim T$, then leads to the corrected Hawking temperature \cite{0106080,2407.21114}
\begin{flalign}\label{TGUP}
T_\text{GUP} &= \frac{Mc^2}{\pi \alpha}\left(1-\sqrt{1-\frac{\alpha \hbar c}{4GM^2}}\right) \\
&= \frac{\hbar c^3}{8\pi G M} + \frac{1}{128} \frac{\hbar^2c^4}{\pi G^2 M^3} + O\left(\alpha^2 \frac{\hbar^3 c^5}{G^3M^5}\right) 
\end{flalign}
GUP (when $\alpha > 0$; we leave the $\alpha<0$ case for later discussion) imposes a minimum uncertainty in the position given by
\begin{equation}\label{xmin}
\Delta x_\text{min} = {\sqrt{\alpha}}{L_{P}}.
\end{equation} 
This is interpreted as implying the presence of a black hole remnant as the end state of Hawking evaporation \cite{0106080}, but whose temperature $T_\text{max}=Mc^2/\pi\alpha$ remains nonzero. As we will discuss in the last section, this nonzero temperature is somewhat problematic, which may be a hint that this expression is not yet the complete form at the Planck scale. 

Utilizing the first law of black hole thermodynamics $dM=TdS$, one can obtain the associated entropy to $T_\text{GUP}$ \cite{0106080}:
\begin{flalign}\label{SGUPfull}
S = ~&2\pi \left(\frac{M}{M_P}\right)^2 + 2\pi \left(\frac{M}{M_P}\right) \sqrt{\left(\frac{M}{M_P}\right)^2 - \frac{\alpha}{4}} \\ \notag &-\frac{\alpha\pi}{2}\ln\left[\left(\frac{M}{M_P}\right)+\sqrt{\left(\frac{M}{M_P}\right)^2-\frac{\alpha}{4}}\right] + C(\alpha),
\end{flalign}
where $M_P$ is the Planck mass and $C(\alpha)$ is a constant term that depends on the (fixed) $\alpha$.
For large $M$ we can expand $S$ into the infinite series
\begin{equation}\label{SGUP}
S = 4\pi \left(\frac{M}{M_P}\right)^2  - \frac{1}{2}\pi\alpha \ln \left(\frac{M}{M_P}\right) + \cdots,
\end{equation}
Henceforth we set $c=1=\hbar$ but we leave $G$ explicit, since we will be discussing effective gravitational constant later. In terms of the area $A$, the coefficient of the logarithmic term would be $-\pi\alpha/4$ since $A\propto M^2$.

The first term in Eq.(\ref{SGUP}) is the usual Bekenstein-Hawking entropy (the area law). The logarithmic correction is expected from various approaches of quantum gravity, including from nonlocal effects \cite{0911.4379,2104.14902,2108.06824}. The higher order terms are expected to be of the form \cite{0002040,1201.6102,Zhang} $\sum_{i=1}^\infty  C_{n+1} (A/G)^{-n}$. These terms are potentially important near the end of the Hawking evaporation. However, since the values and signs of the coefficients $C_{n+1}$ are not completely understood, it is difficult to take these terms into account in general. We will henceforth drop them even though their forms can be obtained explicitly in the GUP case. This is further justified \emph{a posteriori} by the fact that our result (which does not consider these higher order terms) nevertheless recovers the same Hawking temperature as the GUP one. See, however, the last section for what may happen if we do consider these terms.

For practical reasons, one would like to know what the Schwarzschild metric looks like after incorporating GUP. However, it is not clear how such a metric can be obtained. As discussed at length in \cite{2303.10719}, such attempts are typically also heuristic (thus adding even more heuristic arguments upon an already heuristically-derived GUP) and the resulting metric tensors do not agree with each other. It is therefore important to try to look for a more reliable method that can allow us to find the associated metric. For other criticisms of the GUP literature, see \cite{2305.16193}. In this work, instead of considering GUP directly, we aim to find a metric that \emph{reproduces} the GUP-modified Hawking temperature, i.e., Eq.(\ref{TGUP}). The main point is that such a metric is \emph{not} devised in an ad hoc manner just to match the Hawking temperature. We will then explore the consequences of this approach. Notably, the \emph{sign} of the logarithmic correction is found to be opposite to that of Eq.(\ref{SGUP}). Then we comment on the Bekenstein bound in our approach. We will end with more discussions on the caveats and possible extensions, as well as the end state of black hole evaporation.

\section{The Metric From the GEVAG Approach}
We shall will start with the entropy expression
\begin{equation}\label{ent}
S = \frac{A}{4G} + C\ln\left(\frac{A}{G}\right),
\end{equation}
where $C$ is a constant, and employ the equivalence between generalized entropy and varying-$G$ gravity \cite{2407.00484} --- which we refer to as GEVAG for short --- to find the correct associated metric. The idea of GEVAG is easy to state: we know that the Einstein field equations can be derived from thermodynamics via the Jacobson's approach \cite{J}. If we believe that gravity is inherently related to thermodynamics, then any modification to the area law would likewise change the theory of gravity. 

In \cite{2407.00484}, it is shown that if the area law is changed from $S=A/4G$ to $S=F(A)/4G$ for any differentiable function $F$, then assuming that $\nabla^\mu(G_\text{eff}T_{\mu\nu})=0$, the resulting field equation still has the same form as general relativity, with the exception that the gravitational constant $G$ has to be replaced by an effective one $G_\text{eff}=G/F'(A)$. Note the peculiarity that $G_\text{eff}$ is area-dependent except when $S=\eta A/G$ for some constant $\eta$ (unfortunately this does not narrow down to the GR value $\eta=1/4$). The possible interpretations of this result have been discussed at length in \cite{2407.00484}. While this appears quite strange, it is not the first time such an area-dependence has appeared in the literature. In fact, our result remarkably agrees with the asymptotic safe gravity approaches in \cite{2204.09892,2308.16356}, as will be explained shortly. GEVAG also avoids another problem \cite{2407.00484, 2505.03907}: a naive direct application of generalized entropy without taking into account compatibility with the gravity theory generically violate the Bekenstein bound \cite{2207.13652,2411.00694,2405.14799}, but the GEVAG approach does not (possibly up to a constant prefactor). 

Thus, with the logarithmic correction term adding to the area law as in Eq.(\ref{ent}), we have
\begin{equation}
F(A) =A + 4CG \ln\left(\frac{A}{G}\right).
\end{equation}
Therefore, as already discussed in \cite{2407.00484}, GEVAG implies the effective gravitational ``constant'' of the form
\begin{equation}\label{Geff}
G_\text{eff} = \frac{G}{F'(A)}=\frac{G}{1+\dfrac{c_1}{A}}. 
\end{equation}
The constant $c_1$ has physical dimension of area, i.e., $[G^2M^2]$ in our units. The constant prefactor $C$ should have the same physical dimension as $A/G$, and thus $[GM^2]$, which is actually dimensionless in our units\footnote{This is because $[GM_{P}^2]=[G(\hbar c/G)]=1$ in the units $c=\hbar=1$.}. Thus, $C$ is related to $c_1$ by $c_1=4CG$. As pointed out in \cite{2407.00484}, Eq.(\ref{Geff}) agrees with the result in asymptotically safe gravity (ASG)\footnote{Possible connections between ASG and GUP have been explored previously in \cite{2204.07416,2304.00183}.}:
\begin{equation}\label{ASG}
G(k)= \frac{G({k_0})}{1+ \mathfrak{c}k^2},
\end{equation}
where $k_0$ is a reference energy scale and $\mathfrak{c}$ another constant.
if we use the results in ref.\cite{2204.09892} that $k=\text{const.}/\sqrt{A}$. 

In GEVAG, the Schwarzschild solution has the same form as in general relativity:
\begin{flalign}
\d s^2 = &-\left(1-\frac{2G_\text{eff}M}{r}\right)\d t^2 + \left(1-\frac{2G_\text{eff}M}{r}\right)^{-1}\d r^2\\ \notag &+r^2(\d\theta^2+\sin^2\theta \d\phi^2),
\end{flalign}
but with horizon $r_h = 2G_\text{eff} M$ instead of $r=2GM$. 
In this picture, the constant $G$ is still present in the theory, but it is $G_\text{eff}$ that is directly measured in astrophysical observations, in the sense that what we thought of as $GM$ is really $G_\text{eff}M$. In particular, for two black holes of different masses, their $G_\text{eff}$ is actually different. Not knowing this, we would instead deduce different values for their masses. Since the logarithmic correction is minuscule for astrophysical black holes, this effect is quite negligible. For more discussions, including possible cosmological implications, see \cite{2407.00484}.

The relation between $G$ and $G_\text{eff}$ can be solved by considering
the horizon area $A=16\pi G_\text{eff}^2 M^2$. Eq.(\ref{Geff}) then yields 
\begin{equation}
1 = \frac{G(16\pi G_\text{eff} M^2)}{16\pi G_\text{eff}^2 M^2 + c_1}. 
\end{equation}
This reduces to a quadratic equation in $G_\text{eff}$, which is readily solved to give
\begin{equation}
G_\text{eff} = \frac{1}{4}\left(\frac{2\pi GM + \sqrt{4\pi^2 G^2M^2-\pi c_1}}{\pi M}\right).
\end{equation}
We have chosen the plus sign in front of the square root in order to ensure that $G_\text{eff} \to G$ when $c_1 \to 0$.

The Hawking temperature is
\begin{flalign}\label{T}
T &= \frac{1}{8\pi G_\text{eff} M} = \frac{1}{4\pi GM + 2\sqrt{4\pi^2G^2M^2-c_1\pi}} \\ \notag
&=\frac{1}{8\pi GM} + \frac{c_1}{128 \pi^2}\frac{1}{G^3M^3} + O\left(\frac{c^2}{G^5M^5}\right).
\end{flalign}
If we compare with the series expansion of Eq.(\ref{TGUP}) in powers of $\alpha$ (now with $c=\hbar=1$),
\begin{equation}
T_\text{GUP} = \frac{1}{8\pi GM} + \frac{\alpha}{128 \pi}\frac{1}{G^2M^3} + \cdots,
\end{equation}
we see that this has the same form if we identify the GUP parameter $\alpha$ with
\begin{equation}
\alpha = \frac{c_1}{\pi G}=\frac{4C}{\pi}.
\end{equation}
In fact, it is now straightforward to verify that with this identification, Eq.(\ref{T}) agrees \emph{exactly} with Eq.(\ref{TGUP}), not just in the first few orders of correction.

However, a keen-eyed reader would notice that our starting point
\begin{equation}\label{ent}
S = \frac{A}{4G} + C\ln\left(\frac{A}{G}\right) = \frac{1}{4G}\left[{A} + {c_1}\ln\left(\frac{A}{G}\right)\right],
\end{equation}
now has a coefficient $\alpha \pi/4$ in front of the logarithmic term, \emph{not} $-\alpha\pi/4$ as in Eq.(\ref{SGUP}). This discrepancy is due to the use of a different thermodynamic variable. As pointed out in \cite{2407.00484}, in the GEVAG approach, the thermodynamic energy $E$ is no longer the same as the mass $M$. Therefore, in the first law we should use $dE=T dS$ instead. Doing so would lead to
\begin{flalign}\label{energy}
E &= M \sqrt{1-\frac{C}{\pi GM^2}}\\
&=M-\frac{C}{2\pi GM} + O\left(\frac{C^2}{G^2M^2}\right).
\end{flalign}
Up to the first order correction, and again \emph{up to a sign difference}, we observe that such a ``mass correction'' (i.e., the form $M+\text{const.}/M$) had been proposed in the literature. In particular, as pointed out by \cite{2006.04892}, this is consistent with GUP. Essentially, this can be seen from Eq.(\ref{GUP}), since $\Delta x \sim M + 1/M$ if $\Delta p \sim T \sim 1/M$ (neglecting higher order terms). 

In any case, we should emphasize that it is inconsistent to, on one hand, argue that $r_h \sim \Delta x$, and on the other hand, still insist on $r_h = 2GM$ as in the original standard GUP approach, given that $\Delta x \sim M + 1/M + \cdots$. It is therefore interesting to see that the GUP temperature can also be obtained from the GEVAG method, thus avoiding the aforementioned inconsistency.

\section{Bekenstein Bound Analysis}

Now we shall comment on the Bekenstein bound. As was argued in \cite{2505.03907}, the Bekenstein bound for generalized entropy should take the form $S \leqslant C_B RE$, where the ``Bekenstein constant'' $C_B$ need not be $2\pi$. 
The value of $C_B$ can be computed as
\begin{equation}
C_B = \frac{F(A)}{8GG_\text{eff}ME}.
\end{equation}

\begin{figure}[!h]
\centering
\includegraphics[width=0.51\textwidth]{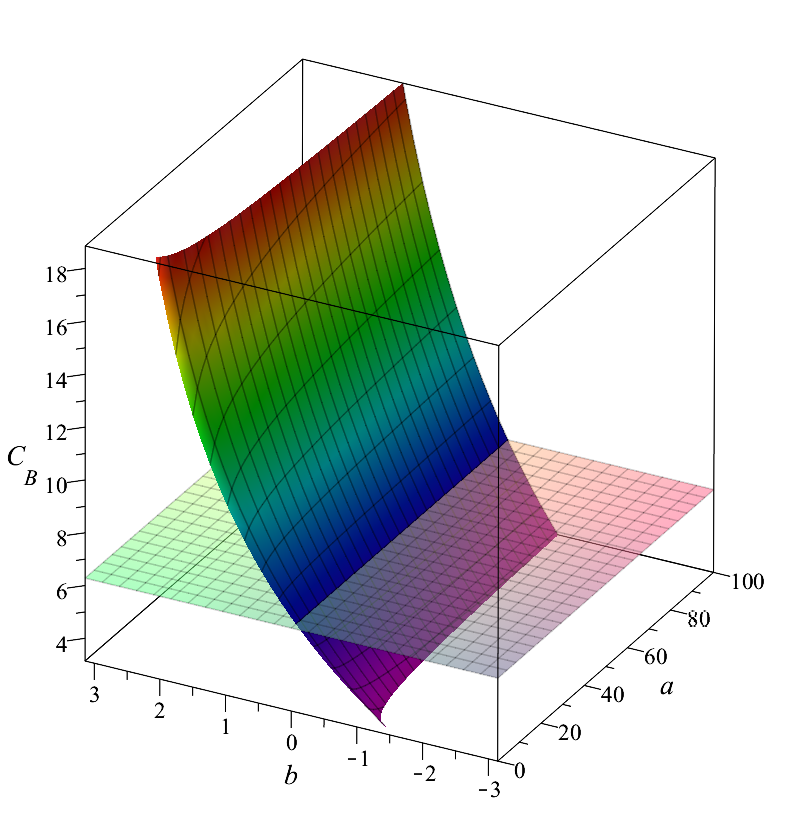}
\caption{The Bekenstein constant $C_B$ as a function of $a \coloneq GM^2$ and $b \coloneq C/a$, where $C$ is the constant in front of the logarithmic correction term of the black hole entropy. We restricted the range to $\pi \leqslant C_B \leqslant 6\pi$ for this plot. The horizontal plane denotes $C_B=2\pi$. We can see that the strict Bekenstein bound is satisfied for $b < 0$, or equivalently, negative GUP parameter. \label{3d}}
\end{figure}

First, let us define the dimensionless quantity 
\begin{equation}
a\coloneq GM^2;
\end{equation}
and purely for convenience, also re-scale $C$ by $a$, writing
\begin{equation}
b \coloneq C/a.
\end{equation}
For the logarithmic correction of the form Eq.(\ref{ent}), $C_B$ has a rather complicated form given by
\begin{equation}
C_B=\frac{\pi^{3/2}\left[2\sqrt{\pi(\pi-b)}+\mathsf{L}(a,b) + 2\pi -b\right]}{\sqrt{\pi-b}\left[\sqrt{\pi(\pi-b)}+\pi\right]},
\end{equation}
where $\mathsf{L}(a,b)$ is a combination of various logarithmic terms:
\begin{equation}
\mathsf{L}(a,b) \coloneq b\left\{\ln a + \ln\left[\frac{\left(2\left(\sqrt{\pi(\pi-b)}+\pi\right)\right)^{2}}{\pi}\right]\right\}.
\end{equation}
The only dependence of $a$ in $C_B$ is via $\ln a$ in $\mathsf{L}(a,b)$.
The plot of $C_B$ as the function of $a$ and $b$ is depicted in Fig.(\ref{3d}).

It can checked that as $b \to 0$ (equivalently $C \to 0$), $C_B$ reduces to the standard value $2\pi$, but $C_B$ is otherwise larger than $2\pi$ if $b>0$. 
In fact, it \emph{diverges} in the limit $b \to \pi$. This corresponds to the case $C=\pi GM^2$, for which the thermodynamic energy given by Eq.(\ref{energy}) vanishes. For a sensible thermodynamical system, we should therefore require that $C \ll \pi GM^2$, i.e., $b \ll \pi$. It is possible that these unwanted pathological behaviors may be artifacts due to the omission of higher order terms in Eq.(\ref{ent}). This is consistent with our claim that the putative remnant with a finite nonzero temperature is problematic. However, such problems do not arise if $b < 0$, or equivalently if $C<0$. We will have more to say about this case soon.

Note that $C_B$ being $a$-dependent does \emph{not} imply that $C$ is mass dependent. In fact, from various approaches to quantum gravity, we expect that $C$ is a constant (albeit model and theory dependent). Thus whatever $C$ is, it should hold also for small black holes. In particular, for Planck mass black hole\footnote{Hereinafter, by ``Planck mass'' black hole we simply mean a black hole that satisfies $a=GM^2=GM_P^2=1$.} (for which $a=1$ and $b=C$), we see that $C \ll \pi$. 

\begin{figure}[!h]
\centering
\includegraphics[width=0.48\textwidth]{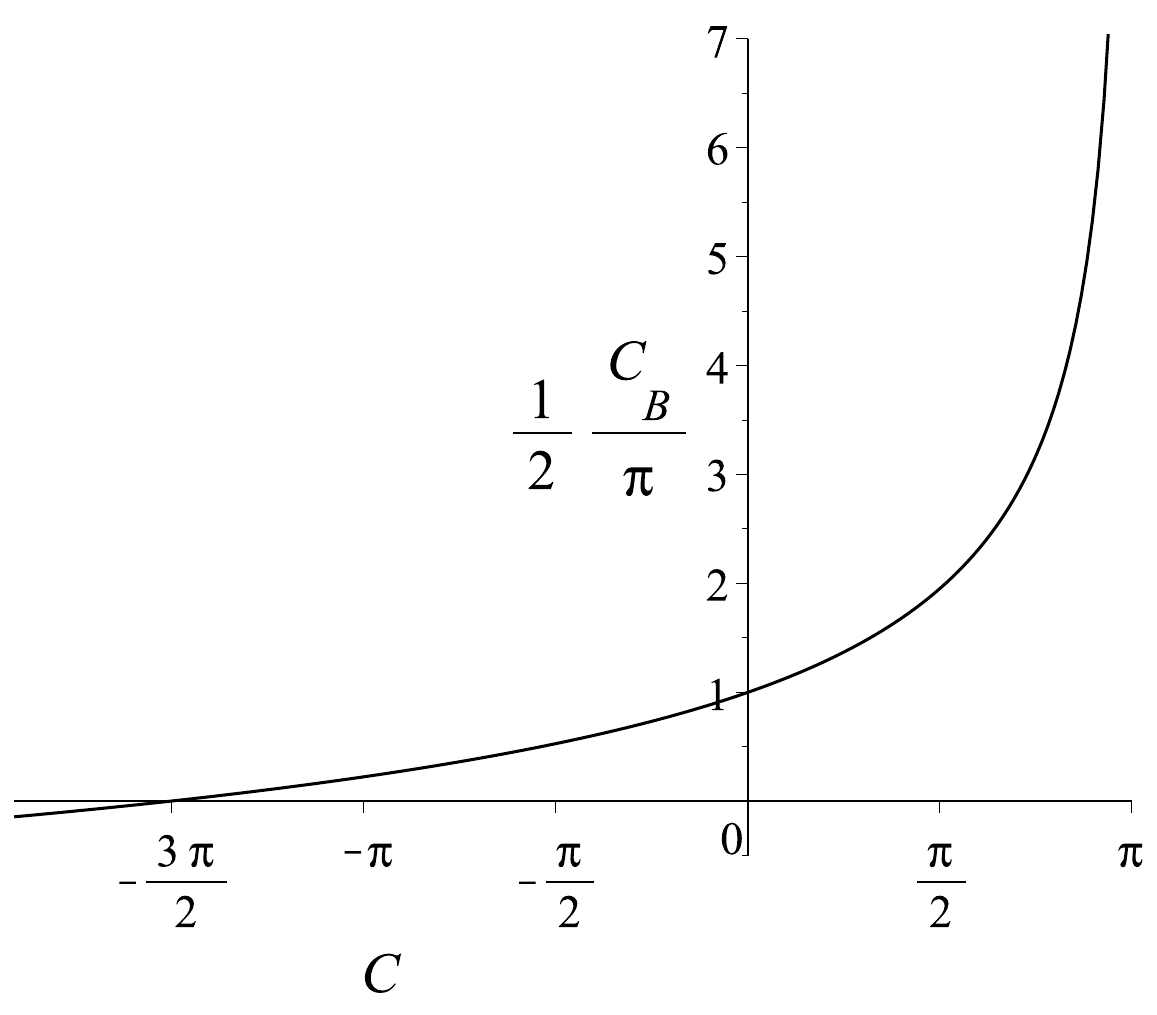}
\caption{The normalized Bekenstein constant $C_B/2\pi$ for Planck mass black hole (that is, whose parameters are $a=1$ and $b=C$). The zero of $C_B$ is close to, but not exactly, $-3\pi/2$. \label{CBplanck}}
\end{figure}

If we want the ``weak Bekenstein bound'' \cite{2505.03907} to hold, such that $C_B/2\pi$ is $O(1)$, we can put constraint on how large $C$ can be. For $b<0$, we have $C_B < 2\pi$, so the Bekenstein bound always holds. However, the magnitude of $C$ should also be constrained from requiring that $C_B>0$. This gives $C \gtrsim -{3\pi}/{2}$ from considering Planck mass black hole, as shown in Fig.(\ref{CBplanck}). 
(The value of $C$ for which $C_B$ vanishes is slightly less than $-{3\pi}/{2}\approx -4.712$, at $k\approx -4.733$.)
This is consistent with the various proposed values for $|C|$ in the literature\footnote{In which the logarithmic correction is usually written in the form $-K \ln A$ for a constant $K>0$. Thus $C=-K$ in our notation.} including $1/2$ and $3/2$ \cite{0002040,0005017,0111001,0210024,0401070}, as well as $\pi/4$ \cite{2203.09968,2407.08358}, $\pi$ or $\pi/2$ (corresponding to $\beta=4\pi$ or $2\pi$ in \cite{ent858}), and even $4 \ln 2$ \cite{2408.02077}. 

One interpretation of the Bekenstein bound is that if a putative system has an entropy that exceeds the bound, it will simply ``restore'' the bound by collapsing into a black hole. Thus if the Bekenstein constant is very small, then black hole formation should become more efficient. The number density of black holes in the universe therefore should also put a lower bound on the Bekenstein constant. From this perspective, just having $C_B \leqslant 2\pi$ is not necessarily good enough, and it may be better to require that $C_B$ does not deviate too much from $2\pi$. This would seem to prefer $C=1/2$ over the other proposed values like $\pi/4$ or $3/2$. However, without seriously conducting a serious study on the number density bound of black holes, no serious conclusion can be drawn from this angle.

The $C<0$ case is interesting because it corresponds to the GUP parameter being negative, which has also been discussed in the literature\footnote{In particular, in \cite{2204.07416} it is argued that the ASG parameter is negative. Although the arguments therein are quite different from the GEVAG approach, if indeed the ASG parameter is negative, this is equivalent to saying that $c_1$, and hence $C$, is negative.} \cite{2204.07416,0912.2253,1804.05176,1806.03691,2201.07919}. In other words, the GEVAG approach implies that if we require the validity of the strict Bekenstein bound, then the GUP parameter has to be negative. 
Note that the conclusion of our approach is exactly the opposite of \cite{2009.12530} --- in their approach, it is the \emph{positive} GUP parameter case that satisfies the strict Bekenstein bound! Again we see a sign difference.

\begin{figure}[!h]
\centering
\includegraphics[width=0.48\textwidth]{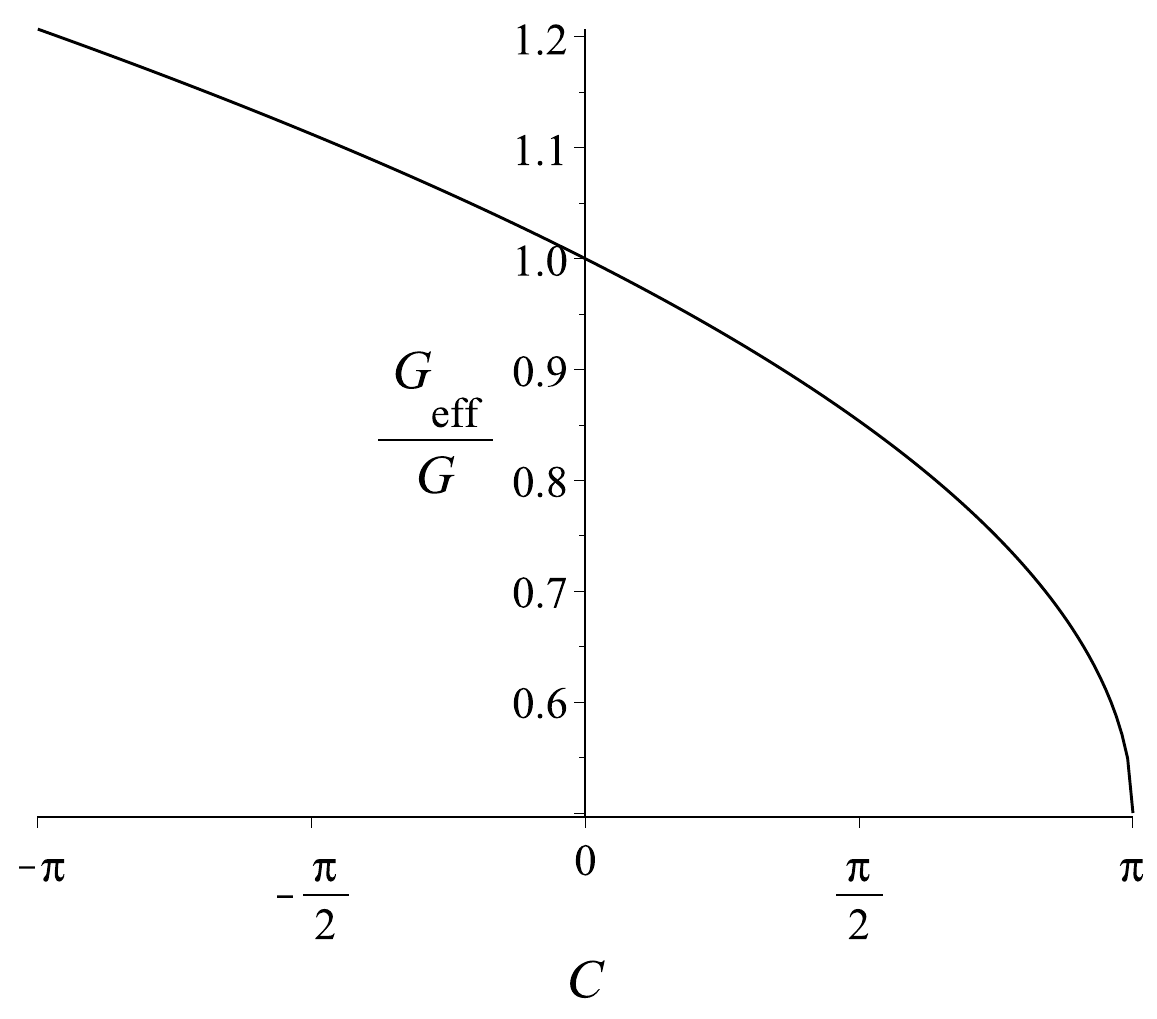}
\caption{The ratio $G_\text{eff}/G$ for a Planck mass black hole $(a=1)$, as a function of $C ~(=b)$. \label{GEFF}}
\end{figure}

Finally, we should remark on $G_\text{eff}$. As can be seen from Eq.(\ref{Geff}), $G_\text{eff}$ is smaller than $G$ if $C>0$, but larger than $G$ if $C<0$. For a Planck mass black hole, $a=1$ and $b=C$, we can re-write Eq.(\ref{Geff}) as\footnote{Coincidentally, we note that $G_\text{eff}/G=(1+\sqrt{5})/2$, the golden section, when $b=-4\pi$.}
\begin{equation}
G_\text{eff}[a=1]=\frac{G}{2}\left(\frac{\pi + \sqrt{\pi(\pi-b)}}{\pi}\right).
\end{equation}
The ratio $G_\text{eff}/G$ is plotted in Fig.(\ref{GEFF}). Mathematically, as $b \to \pi$, we have $G_\text{eff} \to G/2$. Still, as discussed above, one should keep $b$ away from getting too close to $\pi$.

\section{Conclusion \& Discussion: Remnant or Bounce At the Planck Scale?}

In this work, we have used the equivalence between modification of the area law and area-dependent varying-$G$ gravity to derive the Schwarzschild-like black hole solution when the area law receives a logarithmic correction, which is commonly expected from many quantum gravity approaches. (For static solutions, the effective $G_\text{eff}$ is not really varying.) The GEVAG approach is based on the generalization of Jacobson's method that derives general relativity from thermodynamics. If one believes that gravity is inherently related to thermodynamics, then changing the entropy expression would naturally lead to a modified field equation. Using GEVAG to obtain the GUP-corrected metric is therefore more rigorous (though not without problems and challenges \cite{2407.00484}) compared to the various heuristic arguments that have been employed in the literature.
The resulting geometry has Hawking temperature that agrees \emph{exactly} with the temperature expression obtained from the heuristic argument utilizing GUP. In this sense we obtained the ``equivalent GUP metric'' without using any GUP in the derivation\footnote{In other words, the logic here is that from the modified Bekenstein-Hawking entropy, we can derive the modified gravity metric. In comparison, there is another interesting approach in which GUP is argued to be equivalent to metric-affine gravity theory \cite{2311.14066,2401.01159,2401.05941}. On the other hand, there is also an approach to understand GUP from an entropic approach \cite{2206.14166}, but those entropies are not the standard logaritmic correction.}. 
We also showed that the Bekenstein bound is satisfied up to a constant term if the GUP parameter is positive. The original Bekenstein bound holds if the parameter is negative.
While these calculations do lend some credence to the heuristic argument of GUP, there are some differences in our results compared to those of standard GUP. We summarize these in Table \ref{table} at the end of this work. 

One of the peculiarities in the original heuristic approach of GUP is that the horizon was assumed to be unmodified, i.e., $r_h = 2GM$. In fact, as discussed in \cite{2303.10719}, some authors have used this as a guiding principle and proposed that the modified metric function should be of the form \cite{1510.06365,1606.07281}
\begin{equation}
f(r)=\left(1-\frac{2GM}{r}\right)g(r),
\end{equation}
for some suitable $g(r)$. However, there is no fundamental reason why the horizon should be fixed at $r=2GM$. In fact, even in the GUP literature there are arguments that argue for $r\neq 2GM$. For example, in \cite{2006.04892} (see also \cite{1504.07637}), it was proposed that
\begin{equation}\label{carr}
r_h = {2GM}\left[1+\frac{\beta}{2}\left(\frac{M_p}{M}\right)^2\right],
\end{equation}
for some parameter $\beta$. As we argued above, this would be more consistent than assuming $r_h$ to be unmodified.
Indeed in the GEVAG approach, the horizon is no longer the same as the unmodified version, instead we see that
\begin{equation}\label{hor}
r_h = 2G_\text{eff}M = GM + \sqrt{G^2M^2-\frac{\alpha G}{4}}.
\end{equation}
Thus the horizon is smaller than the original Schwarzschild black hole if $\alpha>0$, but larger if $\alpha < 0$. 
In the $\alpha > 0$ case, we observe that
as the Hawking temperature reaches its maximum allowed value, the horizon has shrunk to $r=GM$. This is also when $E$ in Eq.(\ref{energy})becomes zero. Note that Eq.(\ref{hor}), if expanded into power series and keeping only the first order correction, has the same form as Eq.(\ref{carr}). Another observation is that if the horizon is larger, then the entropy is smaller. Likewise, a smaller horizon corresponds to a larger entropy. Although based on a different approach, this was also observed in \cite{1605.06463}.

More importantly, the entropy expression in GEVAG is different from the standard GUP result. The logarithmic correction is $(\alpha\pi/4) \ln (A/G)$ instead of $-(\alpha \pi/4)\ln (A/G)$. This is because in our approach, the thermodynamic energy, the $E$ that should appear in the first law, is no longer the same as $M$ as usually assumed in the GUP approach. 
The opposite \emph{sign} in the two approaches is interesting, as the sign has been discussed from various perspectives in the quantum gravity literature. For example, a negative logarithmic correction is often associated with a microcanonical calculation, while a positive one corresponds to a canonical calculation.
In our approach, we note that perhaps some of these discrepancies may be due to different assumptions on the thermodynamic energy. Likely related to this, is the fact that strict Bekenstein bound is satisfied for negative $\alpha$, which is opposite from the prior result obtained in the literature \cite{0303030,0411022}.
  
A caveat is worth noting: one should also take into account that quantum gravity approaches usually consider higher curvature terms in the gravity theory, so this may be quite different from just general relativity that incorporates $G_\text{eff}$. In other words, the Jacobson's method that assumes the simplest form of Clausius relation $\delta Q = T dS$, may be too restrictive, and generalizations thereof \cite{1112.6215} that aim to include higher curvature terms may be required to investigate this issue in depth.

Putting this aside, let us now comment on the possible end state of black hole under Hawking evaporation. We have seen that $G_\text{eff}$ can be smaller than $G$ if $C$ is positive, but larger if $C$ is negative.
In the former case, when considering a Planck mas black hole, $G_\text{eff}$ never gets smaller than half of the value of $G$. However, for Planck mass black hole,  we should in principle try to investigate what happens if we consider all the higher order corrections in the entropy, with the terms $\sum_{n=1}^\infty  C_{n+1} (A/G)^{-n}$ included. However, as mentioned, the unknown values $C_{n+1}$ present a hurdle towards such an analysis. In any case, the expression for $G_\text{eff}$ would become
\begin{equation}\label{Geffx}
G_\text{eff}=\frac{G}{1+\dfrac{c_1}{A} - \displaystyle\sum_{n=1}^\infty \dfrac{nc_{n+1}}{A^{n+1}}},
\end{equation}
for some dimensionful constants $c_{n+1}$. Since $A$ itself contains $G_\text{eff}$, this equation cannot be solved analytically in general for $G_\text{eff}$ as we did before.
Unfortunately, this is crucial in order to probe what happens at the ``remnant stage'' of GUP, which is rather peculiar. 

As pointed out in a recent review on black hole remnant \cite{2412.00322}, the fact that the remnant still has a nonzero temperature and yet is supposed to be non-radiating may indicate that this state is still not its final form. (The case with negative GUP parameter is different. In that case there is no minimum mass, but the black hole takes an infinite amount of time to evaporate away. Hence there is still an effective remnant \cite{1806.03691}.) The usual argument for remnant \cite{0106080} is that the GUP temperature, Eq.(\ref{TGUP}), should always be real, and so, the black hole stops evaporating when the square root term becomes zero. At this point, however, the temperature is still nonzero. Thus a further perturbation may bring the black hole into a naked singularity or something else entirely. For a comparison, recall that the horizon expression for the Reissner-Nordstr\"om solution also involves a square root. This does not mean that the square root cannot become imaginary; such a state can potentially exist, it just would not be a black hole but a naked singularity. Whether such a violation of the cosmic censorship conjecture is allowed is a nontrivial matter. For our remnant state, which has a nonzero temperature, the risk of instability should be higher than an extremal Reissner-Nordstr\"om black hole with zero temperature. However, this remnant state may be an artifact caused by not taking into account the higher order terms in the entropy correction. Perhaps after taking those terms into consideration, the temperature will reach a maximum and turn towards zero for the Planck size remnant.

Another curious possibility is that the infinite sum in the denominator of Eq.(\ref{Geffx}) may dominate and render $G_\text{eff}$ negative when the black hole becomes sufficiently small, regardless of the sign of $c_1$, and may cause black holes to eventually ``bounce'' back. This is similar to other models that suggested that quantum gravity effects could make gravity effectively \emph{repulsive} at sufficiently high density or energy \cite{1401.6562,Paik,1703.04138,9805060,2412.16989}. Indeed, the latest result from loop quantum gravity, taking backreaction into account, also prefers a ``bounce'' that transits into a white hole instead of a stable remnant \cite{2504.11998}. (See \cite{2407.09584} for a recent review on black-to-white hole transition.)
Similar in spirit, in the Einstein-Cartan-Kibble-Sciama theory of gravity, gravity can become repulsive in the early Universe due to the interactions between the spin of fermionic matter and torsion \cite{1007.0587}. For other approaches, see \cite{1406.7318,2503.14661}. Whether this happens or not in the GEVAG approach would depend on the signs and values of the various $c_{n+1}$'s, which without knowing the full quantum gravity, remain beyond our reach. 
One argument against the change of sign of $G_\text{eff}$ is that by continuity, the denominator in $G_\text{eff}$ will be zero at some point, rendering $G_\text{eff}$ infinite. However, strictly speaking, near the Planck regime, we should not treat black hole area as a continuous parameter, given that we know from various QG approaches that the area is quantized. Therefore, it seems possible that $G_\text{eff}$ can jump from a positive value to a negative one without becoming divergent.

It should be emphasized again that in this work, instead of starting from GUP, we  start from the logarithmic correction of the Bekenstein-Hawking entropy and deduce the effective gravitational constant, and hence the metric, via GEVAG. Nowhere does GUP make any appearance. We connect the result to GUP because the Hawking temperature obtained from this metric matches exactly the Hawking temperature in the standard GUP approach. It would be interesting to further investigate the possible connections between GUP and effective gravitational constant of GEVAG, hopefully in a more direct manner. (In the literature, other approaches also suggested that an effective gravitational constant can arise from GUP \cite{1305.3851,1611.09016}.)

\begin{widetext}
\begin{tabular}{|l||*{5}{c|}}\hline 
\backslashbox{Theories}{Properties}
&\makebox[10em]{Horizon}&\makebox[10em]{Entropy}
&\makebox[12em]{Strict Bekenstein Bound}\\\hline\hline
GUP + GR  & $2GM$& $\dfrac{A}{4G} - \dfrac{\alpha\pi}{4}\ln\left(\dfrac{A}{G}\right) + \displaystyle \sum_{n=1}^{\infty}C_{n+1}\left(\frac{G}{A}\right)^{n}$ & Holds for $\alpha>0$\\ \hline
GEVAG  & $GM + \sqrt{G^2M^2-\dfrac{\alpha G}{4}}$ & $\dfrac{A}{4G} + \dfrac{\alpha\pi}{4}\ln\left(\dfrac{A}{G}\right)$ & Holds for $\alpha<0$\\\hline
\end{tabular}
\captionof{table}{A comparison between the standard GUP with parameter $\alpha$ applied to the Schwarzschild solution in general relativity, and our GEVAG approach. Their Hawking temperature is identical and not specified in the Table. Notice the opposite sign for the logarithmic correction term in the entropy expressions. This perhaps accounts for the opposite sign result for when the strict Bekenstein bound $S \leqslant 2\pi RE$ is satisfied. In GEVAG, the entropy expression is the starting point. Thus we can, in principle, add more higher order correction terms, but the equations cannot be solved analytically even if we know all the coefficients of the higher order terms (which we do not). In that case the temperature will no longer be the same as the GUP one, and the end state may be a remnant or a bounce.  \label{table}}
\end{widetext}

\end{document}